\documentclass[twocolumn,10pt]{revtex4-1}

\usepackage{graphicx,dsfont,color,amsmath,amssymb}

\begin{document}

\title{Generator of arbitrary classical photon statistics}

\author{Ivo Straka}
\email{straka@optics.upol.cz}
\author{Jarom\' ir Mika}
\author{Miroslav Je\v zek}

\affiliation{Department of Optics, Palack\' y University, 17.\ listopadu 1192/12,  771~46 Olomouc, Czech Republic}

\begin{abstract}
We propose and experimentally demonstrate a device for generating light with arbitrary classical photon-number distribution. We use programmable acousto-optical modulation to control the intensity of light within the dynamic range of more than 30 dB and inter-level transitions faster than 500 ns. We propose a universal method that allows high-fidelity generation of user-defined photon statistics. Extremely high precision $<$0.001 can be reached for lower photon numbers, and faithful tail behavior can be reached for very high photon numbers. We demonstrate arbitrary statistics generation for up to 500 photons. The proposed device can produce any classical light statistics with given parameters including Poissonian, super-Poissonian, thermal, and heavy-tailed distributions like log-normal. The presented method can be used to simulate communication channels, calibrate the response of photon-number-resolving detectors, or probe physical phenomena sensitive to photon statistics.
\end{abstract}

\maketitle

\section{Introduction}

Statistics of light is of paramount importance for many applications in metrology, material probing, biomedical optics, and optical communications to name a few. Stellar correlation interferometry represents the first application directly exploiting statistical properties of light to improve the resolution of a star angular diameter measurement \cite{HBT1956}.
In detector metrology, characterization of single-photon detectors or photon-number-resolving detectors is based on probing a detector under test using optical signals with well-defined classical statistics or correlations, and evaluating the corresponding outputs \cite{MigdallBOOK,WalmsleyDetectors,RehacekLoopDetector}.
Light with thermal statistics improves the efficiency of light-matter interaction like optical harmonic generation \cite{Singh1992,Allevi2015,Chekhova2017} and two-photon fluorescence \cite{TwoPhotonFluorescence}. Correlations inherently present in thermal light can enhance interferometric phase sensitivity \cite{SubThermalPhaseResolution} and facilitate sub-wavelength interference \cite{MigdallDoubleSlit}. Ghost imaging is another counterintuitive technique, which requires correlated light signals obtained typically by splitting light with thermal statistics \cite{Boyd2002,Lugiato2004}.
It was also shown that excess optical thermal noise can improve the quantum tomography of an unknown optical signal \cite{NoiseTomography}. Alternatively, a distribution of loss in front of a single photon detector enables to estimate photon statistics of impinging optical signal \cite{Paris2005}. Furthermore, generating arbitrary loss distributions allows us to faithfully simulate fading communication channels \cite{Capraro2012,Usenko2012,Vogel2016PRL,ChannelProbing}.
Asymmetric and heavy-tailed light statistics like log-normal, L\'{e}vy, and Weibull, which play a role in communication through turbulent atmosphere, are also important for optical rogue waves and other rare phenomena attracting a lot of attention recently \cite{Arecchi2016}. The ability to produce light with extreme distributions of optical power or photon numbers would significantly help with simulations of these processes and also probing them.

The statistics of individual photons in an optical signal is a fundamental expression of quantum properties of light.
A semi-classical approach considers the probability density of photon occurrence proportional to the intensity of light \cite{MandelFormula}. Any photon statistics that adheres to this model is called classical. Using this principle, we can stochastically modulate optical intensity to obtain the corresponding photon statistics \cite{TeichModulation}. 
So far, the main focus has been on generating chaotic or pseudo-thermal light, the standard method for which is collecting light scattered by a rotating ground glass \cite{GrainDisc}, a multi-mode fiber \cite{MMthermal}, or other scattering processes \cite{SalehNatPhys2015,SalehOptica2016}. Such an approach is essentially a negative-exponential intensity modulation that results in Bose-Einstein photon statistics.

In this manuscript, we propose a device based on acousto-optical modulation that can be programmed to perform arbitrary intensity modulation. This can be used as a source of light with programmable photon statistics and also for simulation of transmission fluctuations in communication channels. We also propose a new method of obtaining the distribution of optical intensity from an arbitrary photon-number distribution, that is particularly useful for experimental work. We produce several statistics including Poisson, various super-Poisson, thermal, log-normal, bimodal, and uniform distributions. The generated statistics are characterized using a time-multiplexed photon-number-resolving detector and compared with theoretical expectations. Moreover, we demonstrate generating faithful tail behavior of photon statistics and producing highly bunched light. The proposed generator is also readily extensible to the pulse regime or other techniques of modulation.

\section{Photon statistics}

Photon statistics in a light beam of constant power follows a Poisson point process. Let us consider a stochastic variable: the number of photons $n$ in a time window of length $\tau$, while $\mathcal{P}$ is a constant optical power and $E$ the energy of one photon. It follows that the mean number of photons $\langle n \rangle = \mathcal{P}\cdot\tau/E$. Also, $n$ follows the Poisson probability distribution $p_n = e^{-\langle n \rangle}\cdot \langle n \rangle^n/n!$. If the optical power $\mathcal{P}$ is a stochastic variable as well, typically changing in time, we have to consider an integrated optical intensity $W(t) := \int_t^{t+\tau} \mathcal{P}(t)/E\ \mathrm{d}t$, a dimensionless quantity that is governed by a certain probability density $P(W)$. The probability distribution $p_n$ of the number of photons then follows a weighted mixture of Poisson distributions, given by Mandel's formula \cite{MandelFormula}
\begin{equation}
\label{Mandel}
p_n = \int_0^\infty \frac{W^n}{n!} e^{-W}P(W)\ \mathrm{d}W.
\end{equation}

Mandel's formula covers all possible photon statistics for classical states of light. Generating such statistics can be useful if one needs to simulate classical states with no regard to optical phase. Good examples could be found in practical quantum networks -- simulation of optical channels or characterization of single-photon detectors.

\begin{figure}
\centering
\includegraphics[width=\linewidth]{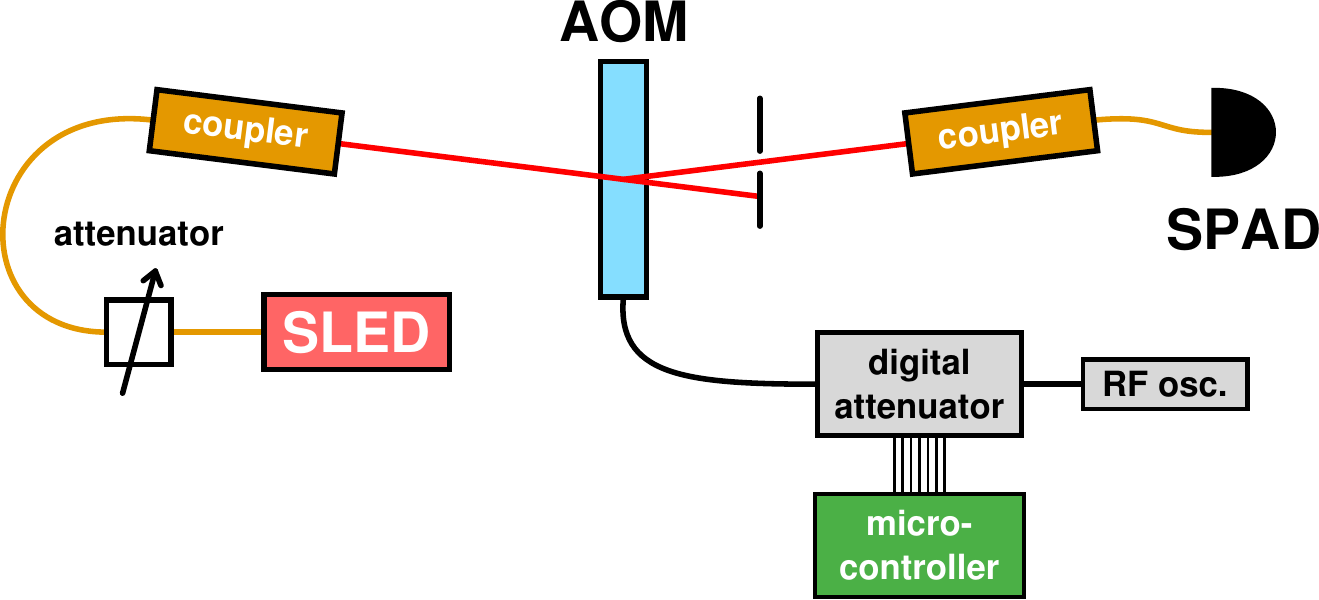}
\caption{The scheme of the generator. Light from a superluminescent diode (SLED) is sent through an acousto-optical modulator (AOM), and the first diffraction order is coupled into a single-mode fiber and detected on a silicon single-photon avalanche diode (SPAD). The AOM is fed from a harmonic signal generator through a programmable attenuator with a parallel 7-bit interface connected to a microcontroller board.}
\label{fig.setup}
\end{figure}

In principle, Eq.\ (\ref{Mandel}) enables engineering photon statistics $p_n$ by modulating $W$ using a corresponding $P(W)$. To this end, we employ an acousto-optical modulator (AOM) driven by a harmonic signal, the amplitude/power of which can be digitally controlled (see Fig.\ \ref{fig.setup}). An attenuated continuous optical beam passes through, while the first diffraction order is collected. The system therefore works effectively as a programmable attenuator with a dynamic range exceeding 30~dB and 128 discrete levels of attenuation. These parameters enable us to span a wide range of $W$ with sufficient sampling to generate a wide variety of user-defined photon statistics, as demonstrated in Results.

\section{Calculating intensity distribution}

If a photon statistics $p_n$ is given instead of the intensity distribution $P(W)$, we need to invert Mandel's formula. There are several different approaches found in the literature \cite{BedardInverse,Perina1969,ByrneInverse,SplinesInverse}, but here we do not need a full inversion to this ill-posed problem. Both spans for $n$ and $W$ are infinite, but covering infinite ranges is experimentally impossible both on detection and generation side. Therefore, we always specify the photon statistics to be generated up to a certain $n_{\text{max}}$. Because we also work with 128 discrete values of intensity $W_i$ in our experiment, Mandel's formula becomes a linear matrix transformation $$\sum_i A_{ni}P_i=p_n$$ with $P_i=P(W_i)$ and $A_{ni} = e^{-W_i}W_i^n/n!$. If we specify the right side, $p_n$, the task is to find the solution vector $P_i$.

Let us discuss the properties and dimensions of the problem. There are two additional constraints, the non-negativity of probabilities $P_i \geq 0$ and summing to unity $\sum_i P_i = 1$. The unity condition can be simply added as an additional equation, extending $A$ and $p$, but the existence of a non-negative solution $P$ is not guaranteed, even though the system of equations is strongly underdetermined: the dimensions of the matrix $A$ extended by the aforementioned row are $(n_{\text{max}}+2) \times 128$, where $n_{\text{max}}$ is set between 10 and 15 for our experiment. This formally leads to a highly multi-dimensional sub-space of solutions, where non-negativity may appear like a weak restriction, but it actually intersects the solution sub-space with one orthant ($1/2^{128}$). Whether or not there is an intersection is neither likely nor trivial to evaluate for high dimensions. In lower dimensions, one can imagine intersecting one octant in 3D space with a plane, line or a point, which depends on their orientation and position and is by no means likely even for a hyperplane. Solution would depend on the choice of $p_n$ and the character of the transformation $A$. We do, however, have a degree of freedom in scaling the intensity levels $W_i$. These are a product of the intensity at the input and the attenuation levels of the AOM system. Although the attenuation levels are fixed, the input power is a free parameter, expressed by the maximum available intensity $W_{\text{max}}$, which is user-scalable and all other intensity levels derive from it. To summarize, our problem has a form

\begin{align*}
\begin{pmatrix}
A_{0,0} & \dots & A_{0, 127} \\
\vdots & \ddots & \vdots \\
A_{n_{\text{max}}, 0} & \dots & A_{n_{\text{max}}, 127} \\
1 & \dots & 1
\end{pmatrix}
&
\begin{pmatrix}
P_0 \\
P_1 \\
\vdots \\
P_{127}
\end{pmatrix}
=
\begin{pmatrix}
p_0 \\
p_1 \\
\vdots \\
p_{n_{\text{max}}} \\
1
\end{pmatrix}
,\\\nonumber
& P_i \geq 0\ \forall i.
\end{align*}

The question now is whether we can be content with finding an approximate solution. In that case, we would have to choose a statistical metric that would define the optimal solution. Later in this work, we elect total-variation distance for quantifying the agreement between model and data, but for the generation itself, we believe a physical argument is needed in order to designate ``the closest photon statistics''. To avoid this ambiguity, we restrict ourselves to precise solutions in this step.

We chose a Python-implemented non-negative least squares (NNLS) algorithm published in Ref.\ \cite{NNLS} as an efficient method of finding such a solution. When defining the photon statistics, we always kept $n_{\text{max}}$ sufficiently low and chose $W_{\text{max}}$ such that the solution would be exact within machine precision. Because the matrix rank is $n_{\text{max}}+2$, we typically end up with the same number of non-zero elements of $P_i$.

If we relax the condition of precise solutions, we can extend $n_{\text{max}}$ at the expense of error. This approach will be investigated in the future.

\section{Experimental implementation}

In our experimental setup, we used a superluminescent diode due to its long-term power stability better than $10^{-4}$ (QPhotonics QSDM-810-2). It was centered around 810 nm and coupled to a single-mode fiber. Light was decoupled into free space and sent through an AOM (Brimrose TEM-125-10-800). The first diffraction order was collected into a single-mode fiber and measured. The AOM was driven by a 125-MHz harmonic signal generated in a harmonic signal generator and passed through a digital step attenuator (Mini-Circuits ZX76-31R75PP+). The attenuator was controlled by an ARM microcontroller (Arduino Due) through a 7-bit parallel interface \cite{Hosak2018}. There were 128 attenuation levels separated by 0.25 dB with a switching speed below 0.5 $\mu$s. The specified responses are 300 ns for the attenuator and 150 ns for the AOM. The diffracted optical intensity was approximately linear with respect to the RF signal power.

For detection, we used a silicon single-photon avalanche diode (SPAD, Excelitas SPCM CD3543H). The width of the detection window $\tau$ was 10 $\mu$s, which is much larger than the recovery time of the SPAD (23 ns), so that the SPAD can distinguish the number of photons incident during the detection window. Data were collected using an electronic time-to-digital converter with a timing resolution of 156 ps (UQDevices UQD-Logic-16).

The intensity modulation was a stepwise sequence of intensity levels and fully controlled by the microcontroller. The sequence was random with a user-specified statistical distribution implemented by an integrated hardware random number generator. The modulation period was 1 ms, which is much longer than the detection window. Together with fast level switching, we can consider the optical intensity in each detection window constant to model the response of the detector. The overall model of the data is then a statistical mixture of constant-intensity models governed by $P_i$, just like the intensity itself.

The detection model has to take into account all relevant imperfections of the detector \cite{MigdallBOOK}. Some have no impact on the measurement. Finite detection efficiency is simply considered as part of the overall attenuation. Background counts are very low and contribute to an offset in intensity, which is accounted for in calibration. However, the effects of recovery time and afterpulsing have a measurable effect on the measured photon statistics. We therefore developed a model, where the detector is inactive for the duration of recovery time after each detection event. Recovery time can be measured directly (23 ns in our case). After recovery time, there is a certain probability of emitting an electronic afterpulse (2.35 \%). The temporal probability distribution of afterpulses was established by a separate measurement. Also, there is a probability that an afterpulse will be emitted right after recovery time (in some literature called a twilight pulse). This probability is proportional to the mean count rate; in our case by a constant $2\cdot 10^{-9}$ s.

All of these effects can be measured, simulated for each intensity $W_i$ and the result compared to measured data. We found that for constant intensities (Poisson light), the measured photon statistics differs from the predicted model by less than $6\cdot 10^{-4}$ for each $p_n$. All of our data that use NNLS inversion have accuracy comparable to this systematic error (for details, see Discussion).

\section{Results}

\begin{figure*}[p]
\centering
\includegraphics[width=\textwidth]{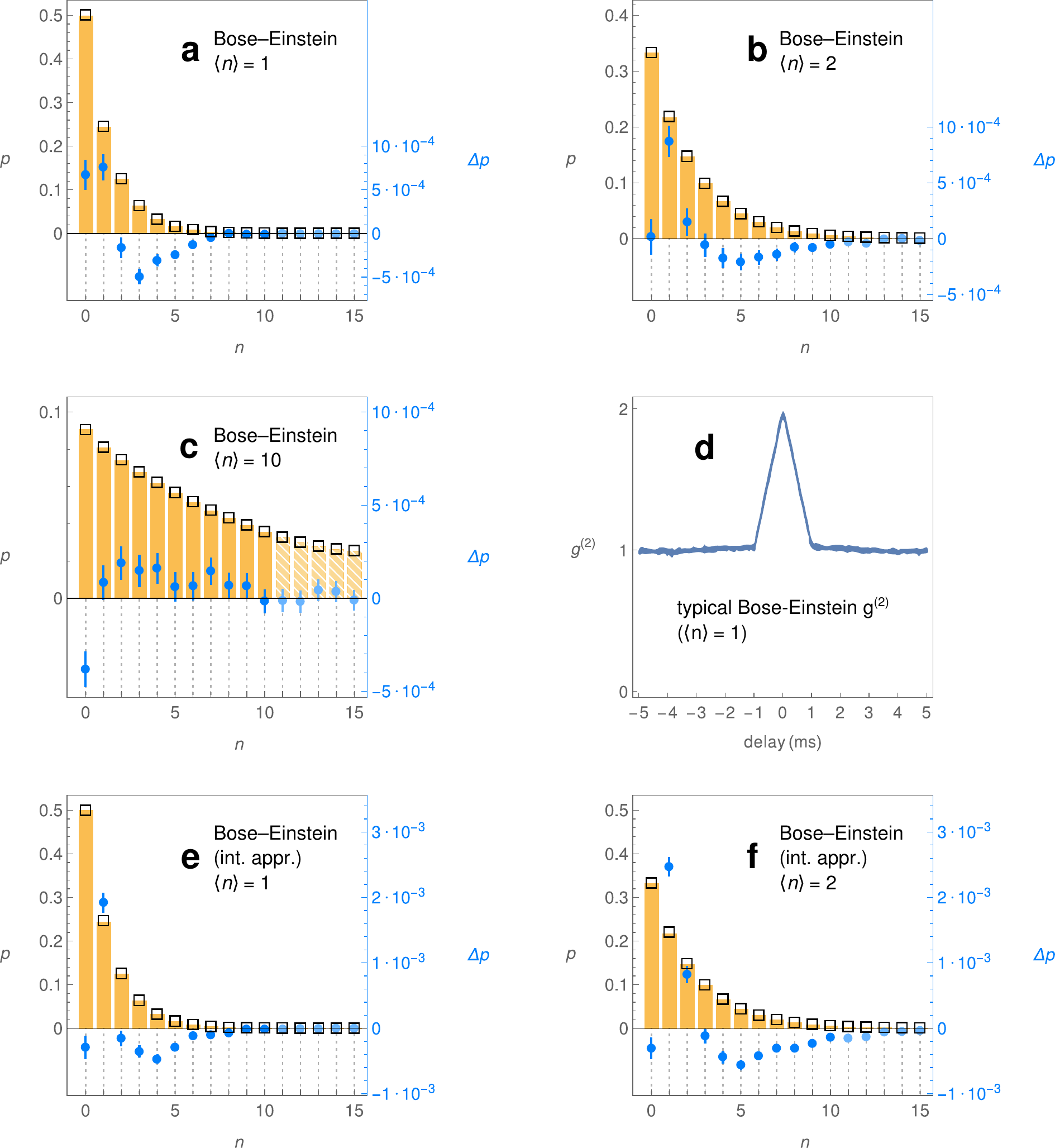}
\caption[Generated Bose-Einstein statistics.]{
Generated Bose-Einstein statistics. Additional information in Table \ref{table.data}. Orange bars represent the expected model. Note that the model is not strictly equal to theoretical $p_n$, because it includes SPAD recovery time and afterpulses. Squares represent measured data. The difference between them, $\delta p = p_{\text{data}}-p_{\text{model}}$, is represented by blue points on a magnified scale. Lighter points and striped bars represent values beyond $n_{\text{max}}$.
\\
\textbf{a -- c:} Bose-Einstein statistics. NNLS was used to calculate $P(W)$.
\\
\textbf{d:} typical shape of the autocorrelation function $g^{(2)}$.
\\
\textbf{e, f:} Bose-Einstein statistics. Intensity was modulated with negative exponential distribution.
}
\label{fig.data1}
\end{figure*}

\begin{figure*}[p]
\centering
\includegraphics[width=\textwidth]{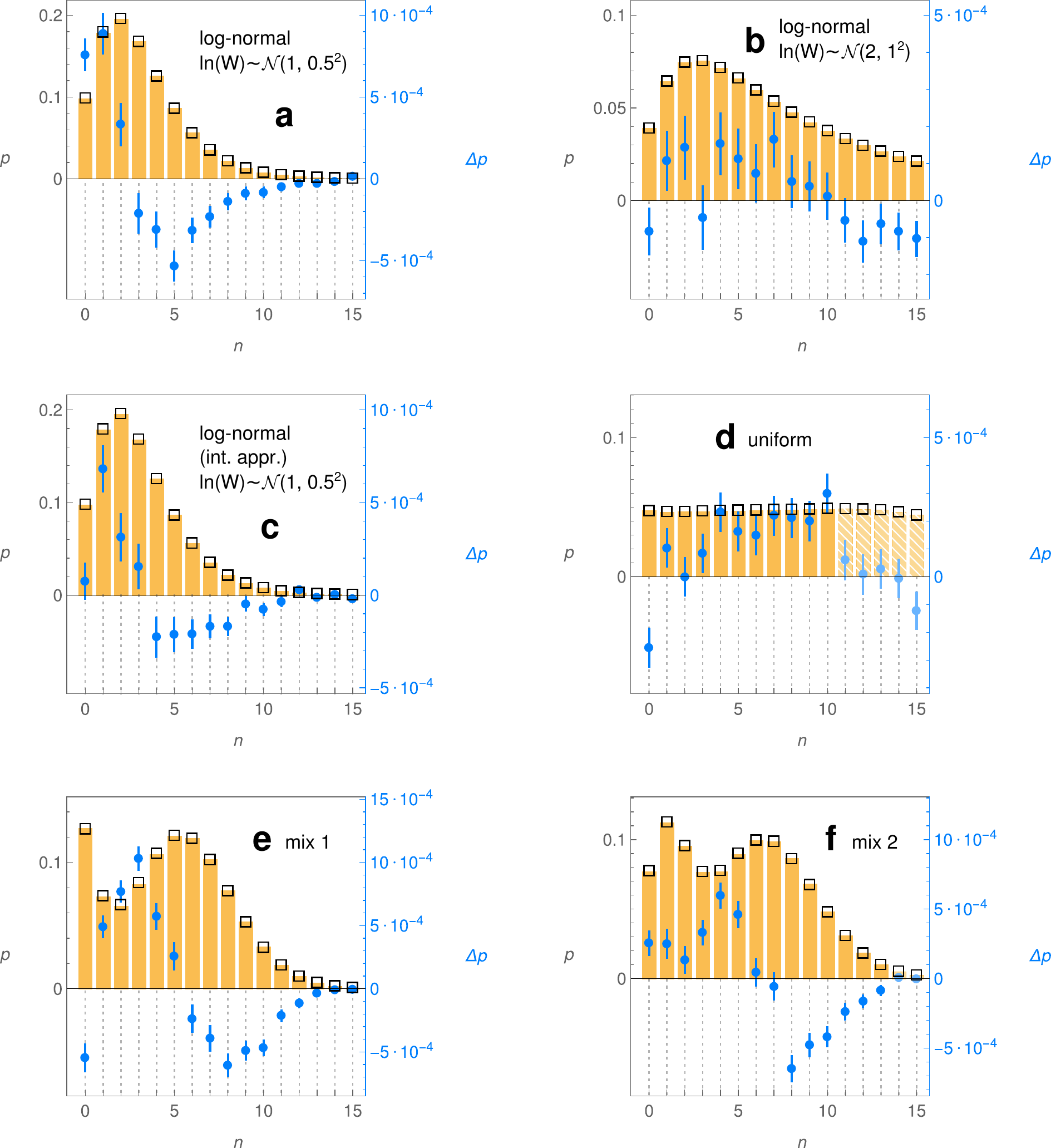}
\caption[Miscellaneous generated photon statistics.]{
Miscellaneous generated photon statistics (continuing from Fig.\ \ref{fig.data1}). Additional information in Table \ref{table.data}. For brevity, we denote a normally distributed variable $X$ with mean $X_0$ and variance $\sigma^2$ as $X \sim \mathcal{N}(X_0,\sigma^2)$.
\\
\textbf{a, b:} desired statistics are based on log-normal intensity distributions. NNLS inversion was used to calculate $P(W)$.
\\
\textbf{c:} same as \textbf{a}, but using log-normal modulation in intensity.
\\
\textbf{d:} a uniform distribution $p = 1/21$. NNLS inversion used.
\\
\textbf{e:} a mixture of 1/4 Bose-Einstein statistics with $\langle n\rangle=1$ and 3/4 normally convoluted Poisson statistics ($W \sim \mathcal{N}(6,0.5^2)$). NNLS inversion used.
\\
\textbf{f: } a mixture of two normally convoluted Poisson statistics, $W \sim 2/3\ \mathcal{N}(1.5,0.25^2)+1/3\ \mathcal{N}(7,0.25^2)$. NNLS inversion used.
}
\label{fig.data2}
\end{figure*}

In Figures \ref{fig.data1} and \ref{fig.data2}, we present various generated photon statistics. We chose Bose-Einstein statistics for its physical significance; it is the statistics of a single mode of light, which is in thermal equilibrium with its environment. The second important statistics we chose is the log-normal. In this case, the log-normal distribution applies to optical intensity. Such light can occur in turbulent optical channels as the result of log-normal fluctuations of transmittance \cite{Capraro2012,Vogel2016PRL}. Finally, we chose some more complex forms of light, bimodal distributions and even a physically bizarre uniform distribution, to demonstrate the variety of statistics that can be realized.

One approach is to take a specified photon statistics $p_n$ and perform the NNLS inversion to obtain an intensity distribution. This inversion-approach is universal and does not require any prior knowledge or decomposition of $P(W)$. The other method (intensity-approach) is simply implementing a given intensity distribution. Here, both limited dynamic range and finite sampling need to be accounted for. The data show that both approaches can lead to accurate results (see Figs.\ \ref{fig.data1}, \ref{fig.data2} and Table \ref{table.data}). For comparison and more detailed considerations, see Discussion.

For the measured data, we need to evaluate the accuracy of the generated photon statistics. We employ a model that arises from the intensity statistics $P(W)$ and models the SPAD response to any intensity $W$. If $P(W)$ was obtained by the inversion-approach, accurate photon statistics is expected only up to $n_{\text{max}}$. Beyond that, high accuracy of the model is achieved if the photons statistics has already been covered enough, $\sum_{n}^{n_{\text{max}}} p_n \to 1$.

To show the difference between model and data, we plot individual differences $\delta p_n$ in Figures \ref{fig.data1} and \ref{fig.data2}. For quantification of the overall difference between the two probability distributions we chose a very conservative definition: total-variation distance $\Delta$. It is defined as the maximum difference between probabilities of any possible set of samples $\{n\}\subseteq \mathds{N}^0$ (for definition details, see Methods). The distances for most photon statistics, and therefore maximal deviations in generated probabilities, are in the order of $\Delta \sim 10^{-3}$ (see Table \ref{table.data}).

The first distribution that we measured was the Bose--Einstein distribution: $p_n = \langle n\rangle^n / (\langle n\rangle+1)^{n+1}$. In theory, it can be generated through negative exponential modulation of the optical intensity $P(W)=\exp(-W/\langle n\rangle)/\langle n\rangle$. For mean values 1 and 2 we successfully employed both approaches. For mean value of 10 and $W_{\text{max}}=20$, the intensity approach fares much worse due to unfavorable scaling of the intensity range required for optimal modulation (see Discussion).

Bose-Einstein distribution is also well-known for its photon bunching, as commonly measured by the autocorrelation function $g^{(2)}$. Its value can be calculated from the photon statistics, but also measured using a Hanbury Brown--Twiss setup \cite{HBT1956}. The data for one such measurement are shown in Fig.\ \ref{fig.data1}(d). We chose a 10-ns coincidence window, although this choice does not influence the result as long as the window is much shorter than the modulation period. The $g^{(2)}$ half-width of 1 ms corresponds to the period of modulation and the measured values at zero for both generation methods are within $1.98 \pm 0.02$ for distributions with $\langle n \rangle=1,2$, which are reasonably well covered by the modulation range. For the mean number 10,  $g^{(2)}$ drops to 1.6, because too much probability for $n$ beyond $n_{\text{max}}$ is not covered in our NNLS inversion.

The second distribution, the log-normal, is given by $P(W)=\exp[-(\ln W - \Omega)^2/(2\sigma^2)]/(\sqrt{2\pi}\sigma W)$, where $\ln(W)$ is normally distributed with mean $\Omega$ and variance $\sigma^2$. The corresponding photon statistics can be numerically calculated by applying Mandel's formula (\ref{Mandel}). Here, again, we used both approaches for mean value $\Omega=1$.

To demonstrate that complex distributions can be generated, as long as they are classical, we chose two concave and one uniform photon statistics. The first was given by a combination of Bose--Einstein statistics and normally convoluted Poisson statistics. The second is a combination of two convoluted Poisson statistics with two distinct peaks. The third one is a uniform distribution with $\langle n\rangle=10$, which is peculiar for its apparent non-classicality. Indeed, if specified in full range, $p_{n \leq 20} = 1/21, p_{n>20} = 0$, the distribution is non-classical. However, if we restrict uniformity only to the first 11 elements, we obtain a classical photon statistics that is partially uniform with a falling tail for $n>10$.

\begin{figure*}
\centering
\includegraphics[width=.7\textwidth]{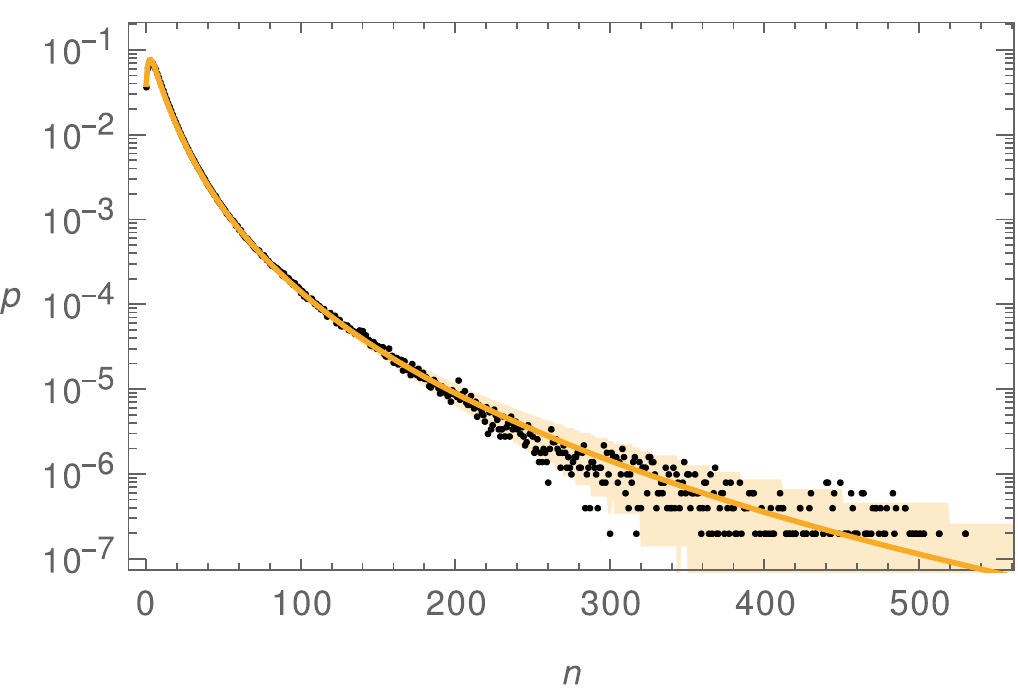}
\caption{Measured log-normal distribution log-$\mathcal{N}(2,1^2)$ with pronounced heavy-tailed behavior. Black points denote experimental data and the orange curve is the theoretical expectation. Orange area denotes statistical confidence region of $2\sigma$, so approximately 95 \% of all data should be within. For this measurement, the modulation period was 2 ms. We also extended the detection window to 200 $\mu$s to avoid detector saturation for high photon numbers. Measurement time was 1000 s.}
\label{fig.lognormal}
\end{figure*}

\begin{figure*}
\centering
\includegraphics[width=.6\textwidth]{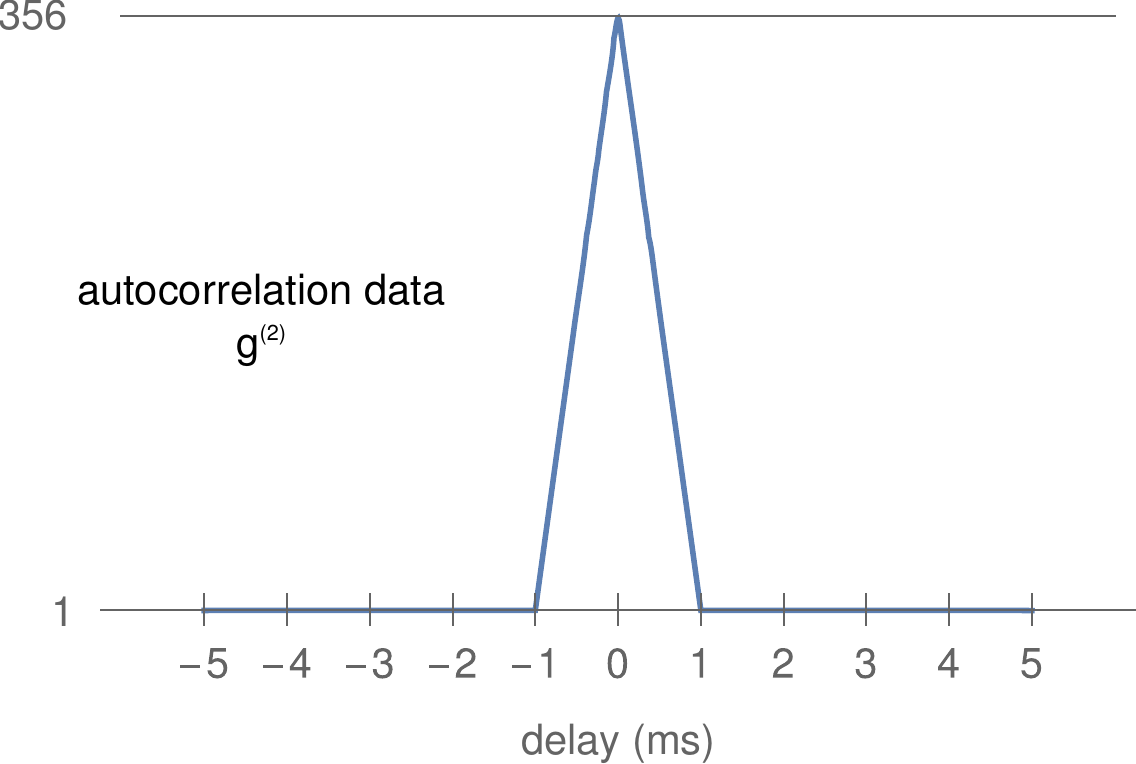}
\caption{The maximum measured autocorrelation value that is achievable using the device in its present form; the result of a two-detector coincidence measurement. Blue curve represents data for a coincidence window of 10 ns, delay values were sampled by 10 $\mu$s and the uncertainty is lower than the thickness of the curve. The shape corresponds to stepwise intensity modulation with 1-ms period. The signal was a random mixture of two intensities with count rates on each detector approximately 3 Mcps (probability $p=6.6\cdot10^{-4}$) and 2 kcps (probability $1-p$). The intensities were chosen such that dark counts and recovery time of the detector would have the smallest effect.}
\label{fig.maxg2}
\end{figure*}

\begin{table}
\centering
\begin{tabular}{rcccc}
\hline
statistics & subj. & $n_{\text{max}}$ & $W_{\text{max}}$ & $\Delta$ [$\times10^{-3}$] \\\hline
B--E(1) & $n$ & 10 & 15 & $1.4$ \\
B--E(1) & $W$ & -- & 13 & $1.9$ \\
B--E(2) & $n$ & 10 & 15 & $1.0$ \\
B--E(2) & $W$ & -- & 20 & $3.3$ \\
B--E(10) & $n$ & 10 & 20 & $0.6$ \\
log-$\mathcal{N}(1,0.5^2)$ & $n$ & 15 & 20 & $2.0$ \\
log-$\mathcal{N}(1,0.5^2)$ & $W$ & -- & 30 & $1.3$ \\
log-$\mathcal{N}(2,1^2)$ & $n$ & 15 & 30 & $1.4$ \\
log-$\mathcal{N}(2,1^2)$ & $W$ & -- & 500\phantom{0} & $14.7$\phantom{0} \\
$\frac{1}{4}\text{B--E}(1)+\frac{3}{4}\mathcal{N}(6,0.5^2)$ & $n$ & 15 & 15 & $3.1$ \\
 $\frac{1}{3}\mathcal{N}(1.5,0.25^2)+\frac{2}{3}\mathcal{N}(7,0.25^2)$ & $n$ & 13 & 15 & $2.1$ \\
uniform$(0,20)$ & $n$ & 10 & 20 & $1.9$ \\
\hline
\end{tabular}
\caption{Results for generated photon statistics. Data shown in Figs. \ref{fig.data1}, \ref{fig.data2} and \ref{fig.lognormal}. Subject denotes the physical quantity specified. For number of photons $n$, NNLS inversion was used to obtain $P(W)$, and for intensity $W$, distribution $P(W)$ was given directly. $n_{\text{max}}$ represents the upper limit on the given probability space and $W_{\text{max}}$ is the maximum intensity. $\Delta$ is the total-variation distance. Definitions of statistical notations follow. B--E($\langle n\rangle$): Bose--Einstein photon statistics or negative exponential intensity distribution, where $\langle n\rangle=\langle W\rangle$. $\mathcal{N}(\langle W\rangle,\sigma^2)$: normally distributed intensity with mean $\langle W\rangle$ and variance $\sigma^2$, or the corresponding photon statistics. log-$\mathcal{N}(\langle\ln W\rangle,\sigma^2)$: log-normal distribution of intensity or the corresponding photon statistics. The moment parameters are the same as for the normal distribution, except here they pertain to $\ln(W)$. uniform($n_1$,$n_2$): uniform photon statistics $p_n=1/(n_2-n_1+1)$ for $n_1 \leq n \leq n_2$.}
\label{table.data}
\end{table}

Finally, Fig.\ \ref{fig.lognormal} shows a log-normal statistics in a scope up to 500 photons with total-variation distance $\Delta = 1.5\cdot 10^{-2}$. Our goal was not to achieve as high precision for all $p_n$ as in previous results, but to demonstrate heavy-tailed scaling of the statistics by comparing measured data to the log-normal. We employed log-normal modulation in intensity and directly compared the measured photon statistics to the theoretical distribution. We did not take into account the detector response, instead, we extended the detection window to avoid too much saturation.

\section{Discussion and methods}

\paragraph{Scope and extensions.} The proposed method works with arbitrary form of intensity modulation. Other modulators like electro-optical, electro-absorption or micromechanical can be used. The advantage of the AOM is its stability, repeatability, high dynamic range and relatively fast response. We estimate that AOMs can reach at least 40 dB of dynamic range and a modulation speed up to several hundred MHz. The repeatability and stability are extremely good and indistinguishable from SLED and coupling stability. To the best of our knowledge, such parameters could not be simultaneously reached using other modulation techniques.

The response speed itself can be further improved by using electro-optical modulation or electro-absorption modulation. Particularly, we successfully tried using an electro-optical integrated Mach-Zehnder amplitude modulator. Compared to acousto-optics, this technology offers very high modulation speeds up to 40 GHz, but typically has only about 20 dB of range and poor long-term stability due to interferometric phase drift. However, an active phase lock can be implemented to solve this issue. That would increase modulation speeds necessary for shorter temporal modes.

In our particular generator, dynamic range is limited by the electronics driving the AOM, which can be in principle improved upon. Generally, dynamic range can be augmented using multiple modulators in a sequence.  This would enable considerable scaling, increasing the range of the generated photon statistics ($n_{\text{max}}$) by orders of magnitude.

\paragraph{Pulse regime.} The proposed scheme would work the same for pulsed light. Wide wavelength spectrum of the pulsed signal does not represent an issue, as the implemented generator uses a superluminescent diode with 20-nm-wide spectrum. On the detection side, a pulsed-domain photon-number-resolving detector would be required.

In this work, we elected the continuous-wave regime to demonstrate the accuracy of the proposed generator. This enabled us to use photon-number resolution in time using one SPAD detector, which is accurate and available. Such an approach offers easily scalable detection range up to hundreds of photons with well-understood imperfections like afterpulsing, recovery time, detection efficiency, and dark counts. These either bear no effect or can be directly measured and taken into account. This detection technique enabled us to demonstrate the quality of generated statistics, but the same quality can be reached in the pulse regime.

\paragraph{Speed and timing.} There are several important timing parameters: recovery time $t_D$, the temporal width $\tau$ of the measurement mode, the period of the modulation $t_M$, and the overall measurement time $T$. They need to be ascending by orders of magnitude, $t_D \ll \tau \ll t_M \ll T$. The recovery time is given by the detector, $t_D=23$ ns. We chose $\tau=10$ $\mu$s so that the SPAD can recognize the number of detections without too much saturation. The period $t_M=1$ ms so that the integral intensity $W(t)$ is influenced only by the given modulation statistics $P(W)$ and does not get distorted by modulation transitions too much. The overall measurement time $T=100$~s to gather enough statistical data. During this time, the fluctuations in the input power were negligible with respect to other systematic errors. For pulsed light and fast photon-number-resolving detector, the sub-500-ns modulation transition limits the possible repetition rate to 2~MHz, provided we desire pulse-to-pulse-independent statistics.

\paragraph{The SPAD model accuracy.} The model of the SPAD assumes only such processes take place in the detector, that depend on the time of only the most recent count event. Apart from recovery time, we included the effect of afterpulses, including their temporal characteristic and their component that grows linearly with count rate \cite{MigdallBOOK}. This turns out to be accurate in the order of $\delta p \sim 10^{-4}$. It is important to note that no data fitting occurred. All parameters were established by an independent measurement beforehand to create a unified detection model. We used continuous Poissonian signals with constant count rates of 50, 100, 200, 500, 1000 and 1500 kcps. The outcomes of the model for various statistics were then compared to the respective raw data.

\paragraph{Systematic errors.} The systematic errors in our data are caused by systematic errors of the SPAD model and by systematic errors in photon statistics on the generation side. For the inversion-approach, generated $p_n$ are precise, because we can measure intensity levels $W_i$ directly. Thus, the systematic error is virtually on the detection side, which is in accord with our measurements. Consequently, we believe the proposed generator could be potentially used for calibrating photon-number-resolving detectors response to various photon statistics.

For the approach, where $P(W)$ is given directly on an infinite scale, finite dynamic range of the modulation and discrete sampling do not permit a precise solution, and both sources of systematic error combine. The errors in the respective data conform to this explanation.

\paragraph{Total-variation distance.} When measuring the number of photons $n$, the set of possible outcomes is $\mathds{N}^0 = \{0, 1, 2, \dots\}$. A photon-number distribution assigns a probability to each individual outcome $n \in \mathds{N}^0$, and by extension to all sets of outcomes $\{n\} \subseteq \mathds{N}^0$. We wish to compare two probability distributions in the broadest possible terms, which means for every set of outcomes. For a certain worst-case set $\{n\}_\Delta$, the difference between probabilities is maximal and thus defined as the total-variation distance $\Delta$. If we know the individual differences $\delta p_n = p_{n,\text{data}}-p_{n,\text{model}}$, it follows from additivity of $p_n$ that the maximum difference can be obtained by summing all the positive or all negative differences. These positive and negative sums are equal, because we know that $\sum_n \delta p_n = 0$ due to probability normalization. Therefore, the total variation distance is a simple sum $\Delta=\frac{1}{2}\sum_n\left| \delta p_n \right|$. We picked this quantification because it is a standard statistical measure, it has a straightforward definition and conservative upper-bound character.

\paragraph{Intensity scaling.} For the intensity-approach, there is a degree of freedom in the scale of $W$ within a constant dynamic range. There is an optimum $W_{\text{max}}$ that gives the best match in $p_n$, and it may vary depending on the definition of statistical distance, but only slightly. The optimal $W_{\text{max}}$ increases with the mean intensity $\langle W\rangle$ much faster than it does for inversion-approach. This is an interesting advantage of the inversion-approach as compared to explicitly given modulation, because it permits photon statistics engineering for intensities where the detector is less saturated. If we use the same $W_{\text{max}}$, the inversion approach gives much better results for higher mean values.

\paragraph{Photon-number correlations.} An important insight is provided by splitting the generated light on a balanced beam splitter and evaluating photon-number correlations between both outputs \cite{Allevi2015}. Examples of measured correlation coefficients are $<$\,0.001 for Poissonian light, $0.48$ for Bose-Einstein light ($0.50$ in theory), and $0.45$ for log-normal light ($0.46$ in theory). The measured values depend on mean number of photons, beam-splitter balance and detection efficiencies. In theory, the correlation parameter for classical light intensity with mean $\mu$ and variance $\sigma^2$ is $C = 1/(1+{2\mu}/{\sigma^2})$, assuming balanced detection.

Another well-known correlation metric is the autocorrelation function $g^{(2)}$ that can be calculated from the photon-number distribution or directly measured using the same two-way-splitting scheme and evaluating coincidence detections. Using aforementioned intensity moments, $g^{(2)}(0) = 1 + (\sigma / \mu)^2$. The highest achievable value stems from dynamic range of the modulation $d = W_\text{max}/W_\text{min}$, and is then $g^{(2)}(0) = (1+d)^2/4d$. It can be generated using a statistical mixture of the minimum and maximum intensity. Using the generator, we were able to reach $g^{(2)}(0) > 350$ (see Fig.\ \ref{fig.maxg2}).

\section{Conclusion}

In conclusion, we presented a generator of arbitrary classical photon statistics than can be fully programmed by the user. We generated various statistics including Poissonian, super-Poissonian, thermal, and log-normal. We demonstrated very high generation accuracy $\delta p_n < 10^{-3}$, which corresponds to the accuracy of the detection mechanism employed. Furthermore, we proposed an efficient inversion method to turn an arbitrary photon statistics into an optical intensity distribution. Finally, we demonstrated high-fidelity generation of photon statistics in the range of 500 photons.

The concept of the generator can be extended to any form of intensity modulation with possible increases in speed and range of generated statistics by orders of magnitude. The generator can also be straightforwardly used in a pulsed regime to produce single-mode states with given statistics. Another use is stochastic loss modulation to simulate realistic transmission channels. As a purely experimental advance, the proposed method of modulation is capable of putting out bunched light with much higher intensity than the conventional rotating glass approach. In addition, the presented generator can produce super-bunched light up to $g^{(2)}(0) \sim 350$.

The presented generator can be used to simulate communication channels, calibrate the response of photon-number-resolving detectors, or probe physical phenomena sensitive to photon statistics.

\section*{Funding}
The Czech Science Foundation (17-26143S).

\section*{Acknowledgments}
This work was supported by the Czech Science Foundation (17-26143S). J. M. also acknowledges the support of Palack\' y University (IGA-PrF-2017-008).

\section*{Disclosures}
The authors declare that there are no conflicts of interest related to this article.

\bibliography{generator_references}

\begin{thebibliography}{31}%
\makeatletter
\providecommand \@ifxundefined [1]{%
 \@ifx{#1\undefined}
}%
\providecommand \@ifnum [1]{%
 \ifnum #1\expandafter \@firstoftwo
 \else \expandafter \@secondoftwo
 \fi
}%
\providecommand \@ifx [1]{%
 \ifx #1\expandafter \@firstoftwo
 \else \expandafter \@secondoftwo
 \fi
}%
\providecommand \natexlab [1]{#1}%
\providecommand \enquote  [1]{``#1''}%
\providecommand \bibnamefont  [1]{#1}%
\providecommand \bibfnamefont [1]{#1}%
\providecommand \citenamefont [1]{#1}%
\providecommand \href@noop [0]{\@secondoftwo}%
\providecommand \href [0]{\begingroup \@sanitize@url \@href}%
\providecommand \@href[1]{\@@startlink{#1}\@@href}%
\providecommand \@@href[1]{\endgroup#1\@@endlink}%
\providecommand \@sanitize@url [0]{\catcode `\\12\catcode `\$12\catcode
  `\&12\catcode `\#12\catcode `\^12\catcode `\_12\catcode `\%12\relax}%
\providecommand \@@startlink[1]{}%
\providecommand \@@endlink[0]{}%
\providecommand \url  [0]{\begingroup\@sanitize@url \@url }%
\providecommand \@url [1]{\endgroup\@href {#1}{\urlprefix }}%
\providecommand \urlprefix  [0]{URL }%
\providecommand \Eprint [0]{\href }%
\providecommand \doibase [0]{http://dx.doi.org/}%
\providecommand \selectlanguage [0]{\@gobble}%
\providecommand \bibinfo  [0]{\@secondoftwo}%
\providecommand \bibfield  [0]{\@secondoftwo}%
\providecommand \translation [1]{[#1]}%
\providecommand \BibitemOpen [0]{}%
\providecommand \bibitemStop [0]{}%
\providecommand \bibitemNoStop [0]{.\EOS\space}%
\providecommand \EOS [0]{\spacefactor3000\relax}%
\providecommand \BibitemShut  [1]{\csname bibitem#1\endcsname}%
\let\auto@bib@innerbib\@empty
\bibitem [{\citenamefont {Hanbury~Brown}\ and\ \citenamefont
  {Twiss}(1956)}]{HBT1956}%
  \BibitemOpen
  \bibfield  {author} {\bibinfo {author} {\bibfnamefont {R.}~\bibnamefont
  {Hanbury~Brown}}\ and\ \bibinfo {author} {\bibfnamefont {R.~Q.}\ \bibnamefont
  {Twiss}},\ }\href {http://dx.doi.org/10.1038/1781046a0} {\bibfield  {journal}
  {\bibinfo  {journal} {Nature}\ }\textbf {\bibinfo {volume} {178}},\ \bibinfo
  {pages} {1046} (\bibinfo {year} {1956})}\BibitemShut {NoStop}%
\bibitem [{\citenamefont {Migdall}\ \emph {et~al.}(2013)\citenamefont
  {Migdall}, \citenamefont {Polyakov}, \citenamefont {Fan},\ and\ \citenamefont
  {Bienfang}}]{MigdallBOOK}%
  \BibitemOpen
  \bibfield  {author} {\bibinfo {author} {\bibfnamefont {A.}~\bibnamefont
  {Migdall}}, \bibinfo {author} {\bibfnamefont {S.~V.}\ \bibnamefont
  {Polyakov}}, \bibinfo {author} {\bibfnamefont {J.}~\bibnamefont {Fan}}, \
  and\ \bibinfo {author} {\bibfnamefont {J.}~\bibnamefont {Bienfang}},\
  }\href@noop {} {\emph {\bibinfo {title} {Single-Photon Generation and
  Detection}}}\ (\bibinfo  {publisher} {Academic Press},\ \bibinfo {year}
  {2013})\BibitemShut {NoStop}%
\bibitem [{\citenamefont {Lundeen}\ \emph {et~al.}(2008)\citenamefont
  {Lundeen}, \citenamefont {Feito}, \citenamefont {Coldenstrodt-Ronge},
  \citenamefont {Pregnell}, \citenamefont {Silberhorn}, \citenamefont {Ralph},
  \citenamefont {Eisert}, \citenamefont {Plenio},\ and\ \citenamefont
  {Walmsley}}]{WalmsleyDetectors}%
  \BibitemOpen
  \bibfield  {author} {\bibinfo {author} {\bibfnamefont {J.~S.}\ \bibnamefont
  {Lundeen}}, \bibinfo {author} {\bibfnamefont {A.}~\bibnamefont {Feito}},
  \bibinfo {author} {\bibfnamefont {H.}~\bibnamefont {Coldenstrodt-Ronge}},
  \bibinfo {author} {\bibfnamefont {K.~L.}\ \bibnamefont {Pregnell}}, \bibinfo
  {author} {\bibfnamefont {C.}~\bibnamefont {Silberhorn}}, \bibinfo {author}
  {\bibfnamefont {T.~C.}\ \bibnamefont {Ralph}}, \bibinfo {author}
  {\bibfnamefont {J.}~\bibnamefont {Eisert}}, \bibinfo {author} {\bibfnamefont
  {M.~B.}\ \bibnamefont {Plenio}}, \ and\ \bibinfo {author} {\bibfnamefont
  {I.~A.}\ \bibnamefont {Walmsley}},\ }\href
  {http://dx.doi.org/10.1038/nphys1133} {\bibfield  {journal} {\bibinfo
  {journal} {Nat. Phys.}\ }\textbf {\bibinfo {volume} {5}},\ \bibinfo {pages}
  {27} (\bibinfo {year} {2008})}\BibitemShut {NoStop}%
\bibitem [{\citenamefont {\ifmmode \check{R}\else
  \v{R}\fi{}eh\'a\ifmmode~\check{c}\else \v{c}\fi{}ek}\ \emph
  {et~al.}(2003)\citenamefont {\ifmmode \check{R}\else
  \v{R}\fi{}eh\'a\ifmmode~\check{c}\else \v{c}\fi{}ek}, \citenamefont {Hradil},
  \citenamefont {Haderka}, \citenamefont {Pe\ifmmode~\check{r}\else
  \v{r}\fi{}ina},\ and\ \citenamefont {Hamar}}]{RehacekLoopDetector}%
  \BibitemOpen
  \bibfield  {author} {\bibinfo {author} {\bibfnamefont {J.}~\bibnamefont
  {\ifmmode \check{R}\else \v{R}\fi{}eh\'a\ifmmode~\check{c}\else
  \v{c}\fi{}ek}}, \bibinfo {author} {\bibfnamefont {Z.}~\bibnamefont {Hradil}},
  \bibinfo {author} {\bibfnamefont {O.}~\bibnamefont {Haderka}}, \bibinfo
  {author} {\bibfnamefont {J.}~\bibnamefont {Pe\ifmmode~\check{r}\else
  \v{r}\fi{}ina}}, \ and\ \bibinfo {author} {\bibfnamefont {M.}~\bibnamefont
  {Hamar}},\ }\href {\doibase 10.1103/PhysRevA.67.061801} {\bibfield  {journal}
  {\bibinfo  {journal} {Phys. Rev. A}\ }\textbf {\bibinfo {volume} {67}},\
  \bibinfo {pages} {061801} (\bibinfo {year} {2003})}\BibitemShut {NoStop}%
\bibitem [{\citenamefont {Qu}\ and\ \citenamefont {Singh}(1992)}]{Singh1992}%
  \BibitemOpen
  \bibfield  {author} {\bibinfo {author} {\bibfnamefont {Y.}~\bibnamefont
  {Qu}}\ and\ \bibinfo {author} {\bibfnamefont {S.}~\bibnamefont {Singh}},\
  }\href {\doibase https://doi.org/10.1016/0030-4018(92)90339-S} {\bibfield
  {journal} {\bibinfo  {journal} {Opt. Commun.}\ }\textbf {\bibinfo {volume}
  {90}},\ \bibinfo {pages} {111} (\bibinfo {year} {1992})}\BibitemShut
  {NoStop}%
\bibitem [{\citenamefont {Allevi}\ and\ \citenamefont
  {Bondani}(2015)}]{Allevi2015}%
  \BibitemOpen
  \bibfield  {author} {\bibinfo {author} {\bibfnamefont {A.}~\bibnamefont
  {Allevi}}\ and\ \bibinfo {author} {\bibfnamefont {M.}~\bibnamefont
  {Bondani}},\ }\href {\doibase 10.1364/OL.40.003089} {\bibfield  {journal}
  {\bibinfo  {journal} {Opt. Lett.}\ }\textbf {\bibinfo {volume} {40}},\
  \bibinfo {pages} {3089} (\bibinfo {year} {2015})}\BibitemShut {NoStop}%
\bibitem [{\citenamefont {Spasibko}\ \emph {et~al.}(2017)\citenamefont
  {Spasibko}, \citenamefont {Kopylov}, \citenamefont {Krutyanskiy},
  \citenamefont {Murzina}, \citenamefont {Leuchs},\ and\ \citenamefont
  {Chekhova}}]{Chekhova2017}%
  \BibitemOpen
  \bibfield  {author} {\bibinfo {author} {\bibfnamefont {K.~Y.}\ \bibnamefont
  {Spasibko}}, \bibinfo {author} {\bibfnamefont {D.~A.}\ \bibnamefont
  {Kopylov}}, \bibinfo {author} {\bibfnamefont {V.~L.}\ \bibnamefont
  {Krutyanskiy}}, \bibinfo {author} {\bibfnamefont {T.~V.}\ \bibnamefont
  {Murzina}}, \bibinfo {author} {\bibfnamefont {G.}~\bibnamefont {Leuchs}}, \
  and\ \bibinfo {author} {\bibfnamefont {M.~V.}\ \bibnamefont {Chekhova}},\
  }\href {\doibase 10.1103/PhysRevLett.119.223603} {\bibfield  {journal}
  {\bibinfo  {journal} {Phys. Rev. Lett.}\ }\textbf {\bibinfo {volume} {119}},\
  \bibinfo {pages} {223603} (\bibinfo {year} {2017})}\BibitemShut {NoStop}%
\bibitem [{\citenamefont {Jechow}\ \emph {et~al.}(2013)\citenamefont {Jechow},
  \citenamefont {Seefeldt}, \citenamefont {Kurzke}, \citenamefont {Heuer},\
  and\ \citenamefont {Menzel}}]{TwoPhotonFluorescence}%
  \BibitemOpen
  \bibfield  {author} {\bibinfo {author} {\bibfnamefont {A.}~\bibnamefont
  {Jechow}}, \bibinfo {author} {\bibfnamefont {M.}~\bibnamefont {Seefeldt}},
  \bibinfo {author} {\bibfnamefont {H.}~\bibnamefont {Kurzke}}, \bibinfo
  {author} {\bibfnamefont {A.}~\bibnamefont {Heuer}}, \ and\ \bibinfo {author}
  {\bibfnamefont {R.}~\bibnamefont {Menzel}},\ }\href
  {http://dx.doi.org/10.1038/nphoton.2013.271} {\bibfield  {journal} {\bibinfo
  {journal} {Nature Photonics}\ }\textbf {\bibinfo {volume} {7}},\ \bibinfo
  {pages} {973} (\bibinfo {year} {2013})}\BibitemShut {NoStop}%
\bibitem [{\citenamefont {Rafsanjani}\ \emph {et~al.}(2017)\citenamefont
  {Rafsanjani}, \citenamefont {Mirhosseini}, \citenamefont
  {Maga{\~{n}}a-Loaiza}, \citenamefont {Gard}, \citenamefont {Birrittella},
  \citenamefont {Koltenbah}, \citenamefont {Parazzoli}, \citenamefont {Capron},
  \citenamefont {Gerry}, \citenamefont {Dowling},\ and\ \citenamefont
  {Boyd}}]{SubThermalPhaseResolution}%
  \BibitemOpen
  \bibfield  {author} {\bibinfo {author} {\bibfnamefont {S.~M.~H.}\
  \bibnamefont {Rafsanjani}}, \bibinfo {author} {\bibfnamefont
  {M.}~\bibnamefont {Mirhosseini}}, \bibinfo {author} {\bibfnamefont {O.~S.}\
  \bibnamefont {Maga{\~{n}}a-Loaiza}}, \bibinfo {author} {\bibfnamefont
  {B.~T.}\ \bibnamefont {Gard}}, \bibinfo {author} {\bibfnamefont
  {R.}~\bibnamefont {Birrittella}}, \bibinfo {author} {\bibfnamefont {B.~E.}\
  \bibnamefont {Koltenbah}}, \bibinfo {author} {\bibfnamefont {C.~G.}\
  \bibnamefont {Parazzoli}}, \bibinfo {author} {\bibfnamefont {B.~A.}\
  \bibnamefont {Capron}}, \bibinfo {author} {\bibfnamefont {C.~C.}\
  \bibnamefont {Gerry}}, \bibinfo {author} {\bibfnamefont {J.~P.}\ \bibnamefont
  {Dowling}}, \ and\ \bibinfo {author} {\bibfnamefont {R.~W.}\ \bibnamefont
  {Boyd}},\ }\href {\doibase 10.1364/OPTICA.4.000487} {\bibfield  {journal}
  {\bibinfo  {journal} {Optica}\ }\textbf {\bibinfo {volume} {4}},\ \bibinfo
  {pages} {487} (\bibinfo {year} {2017})}\BibitemShut {NoStop}%
\bibitem [{\citenamefont {Zhai}\ \emph {et~al.}(2014)\citenamefont {Zhai},
  \citenamefont {Becerra}, \citenamefont {Fan},\ and\ \citenamefont
  {Migdall}}]{MigdallDoubleSlit}%
  \BibitemOpen
  \bibfield  {author} {\bibinfo {author} {\bibfnamefont {Y.}~\bibnamefont
  {Zhai}}, \bibinfo {author} {\bibfnamefont {F.~E.}\ \bibnamefont {Becerra}},
  \bibinfo {author} {\bibfnamefont {J.}~\bibnamefont {Fan}}, \ and\ \bibinfo
  {author} {\bibfnamefont {A.}~\bibnamefont {Migdall}},\ }\href {\doibase
  10.1063/1.4895101} {\bibfield  {journal} {\bibinfo  {journal} {Applied
  Physics Letters}\ }\textbf {\bibinfo {volume} {105}},\ \bibinfo {pages}
  {101104} (\bibinfo {year} {2014})}\BibitemShut {NoStop}%
\bibitem [{\citenamefont {Bennink}\ \emph {et~al.}(2002)\citenamefont
  {Bennink}, \citenamefont {Bentley},\ and\ \citenamefont {Boyd}}]{Boyd2002}%
  \BibitemOpen
  \bibfield  {author} {\bibinfo {author} {\bibfnamefont {R.~S.}\ \bibnamefont
  {Bennink}}, \bibinfo {author} {\bibfnamefont {S.~J.}\ \bibnamefont
  {Bentley}}, \ and\ \bibinfo {author} {\bibfnamefont {R.~W.}\ \bibnamefont
  {Boyd}},\ }\href {\doibase 10.1103/PhysRevLett.89.113601} {\bibfield
  {journal} {\bibinfo  {journal} {Phys. Rev. Lett.}\ }\textbf {\bibinfo
  {volume} {89}},\ \bibinfo {pages} {113601} (\bibinfo {year}
  {2002})}\BibitemShut {NoStop}%
\bibitem [{\citenamefont {Gatti}\ \emph {et~al.}(2004)\citenamefont {Gatti},
  \citenamefont {Brambilla}, \citenamefont {Bache},\ and\ \citenamefont
  {Lugiato}}]{Lugiato2004}%
  \BibitemOpen
  \bibfield  {author} {\bibinfo {author} {\bibfnamefont {A.}~\bibnamefont
  {Gatti}}, \bibinfo {author} {\bibfnamefont {E.}~\bibnamefont {Brambilla}},
  \bibinfo {author} {\bibfnamefont {M.}~\bibnamefont {Bache}}, \ and\ \bibinfo
  {author} {\bibfnamefont {L.~A.}\ \bibnamefont {Lugiato}},\ }\href {\doibase
  10.1103/PhysRevLett.93.093602} {\bibfield  {journal} {\bibinfo  {journal}
  {Phys. Rev. Lett.}\ }\textbf {\bibinfo {volume} {93}},\ \bibinfo {pages}
  {093602} (\bibinfo {year} {2004})}\BibitemShut {NoStop}%
\bibitem [{\citenamefont {Harder}\ \emph {et~al.}(2014)\citenamefont {Harder},
  \citenamefont {Mogilevtsev}, \citenamefont {Korolkova},\ and\ \citenamefont
  {Silberhorn}}]{NoiseTomography}%
  \BibitemOpen
  \bibfield  {author} {\bibinfo {author} {\bibfnamefont {G.}~\bibnamefont
  {Harder}}, \bibinfo {author} {\bibfnamefont {D.}~\bibnamefont {Mogilevtsev}},
  \bibinfo {author} {\bibfnamefont {N.}~\bibnamefont {Korolkova}}, \ and\
  \bibinfo {author} {\bibfnamefont {C.}~\bibnamefont {Silberhorn}},\ }\href
  {\doibase 10.1103/PhysRevLett.113.070403} {\bibfield  {journal} {\bibinfo
  {journal} {Phys. Rev. Lett.}\ }\textbf {\bibinfo {volume} {113}},\ \bibinfo
  {pages} {070403} (\bibinfo {year} {2014})}\BibitemShut {NoStop}%
\bibitem [{\citenamefont {Zambra}\ \emph {et~al.}(2005)\citenamefont {Zambra},
  \citenamefont {Andreoni}, \citenamefont {Bondani}, \citenamefont {Gramegna},
  \citenamefont {Genovese}, \citenamefont {Brida}, \citenamefont {Rossi},\ and\
  \citenamefont {Paris}}]{Paris2005}%
  \BibitemOpen
  \bibfield  {author} {\bibinfo {author} {\bibfnamefont {G.}~\bibnamefont
  {Zambra}}, \bibinfo {author} {\bibfnamefont {A.}~\bibnamefont {Andreoni}},
  \bibinfo {author} {\bibfnamefont {M.}~\bibnamefont {Bondani}}, \bibinfo
  {author} {\bibfnamefont {M.}~\bibnamefont {Gramegna}}, \bibinfo {author}
  {\bibfnamefont {M.}~\bibnamefont {Genovese}}, \bibinfo {author}
  {\bibfnamefont {G.}~\bibnamefont {Brida}}, \bibinfo {author} {\bibfnamefont
  {A.}~\bibnamefont {Rossi}}, \ and\ \bibinfo {author} {\bibfnamefont
  {M.~G.~A.}\ \bibnamefont {Paris}},\ }\href {\doibase
  10.1103/PhysRevLett.95.063602} {\bibfield  {journal} {\bibinfo  {journal}
  {Phys. Rev. Lett.}\ }\textbf {\bibinfo {volume} {95}},\ \bibinfo {pages}
  {063602} (\bibinfo {year} {2005})}\BibitemShut {NoStop}%
\bibitem [{\citenamefont {Capraro}\ \emph {et~al.}(2012)\citenamefont
  {Capraro}, \citenamefont {Tomaello}, \citenamefont {Dall'Arche},
  \citenamefont {Gerlin}, \citenamefont {Ursin}, \citenamefont {Vallone},\ and\
  \citenamefont {Villoresi}}]{Capraro2012}%
  \BibitemOpen
  \bibfield  {author} {\bibinfo {author} {\bibfnamefont {I.}~\bibnamefont
  {Capraro}}, \bibinfo {author} {\bibfnamefont {A.}~\bibnamefont {Tomaello}},
  \bibinfo {author} {\bibfnamefont {A.}~\bibnamefont {Dall'Arche}}, \bibinfo
  {author} {\bibfnamefont {F.}~\bibnamefont {Gerlin}}, \bibinfo {author}
  {\bibfnamefont {R.}~\bibnamefont {Ursin}}, \bibinfo {author} {\bibfnamefont
  {G.}~\bibnamefont {Vallone}}, \ and\ \bibinfo {author} {\bibfnamefont
  {P.}~\bibnamefont {Villoresi}},\ }\href {\doibase
  10.1103/PhysRevLett.109.200502} {\bibfield  {journal} {\bibinfo  {journal}
  {Phys. Rev. Lett.}\ }\textbf {\bibinfo {volume} {109}},\ \bibinfo {pages}
  {200502} (\bibinfo {year} {2012})}\BibitemShut {NoStop}%
\bibitem [{\citenamefont {Usenko}\ \emph {et~al.}(2012)\citenamefont {Usenko},
  \citenamefont {Heim}, \citenamefont {Peuntinger}, \citenamefont {Wittmann},
  \citenamefont {Marquardt}, \citenamefont {Leuchs},\ and\ \citenamefont
  {Filip}}]{Usenko2012}%
  \BibitemOpen
  \bibfield  {author} {\bibinfo {author} {\bibfnamefont {V.~C.}\ \bibnamefont
  {Usenko}}, \bibinfo {author} {\bibfnamefont {B.}~\bibnamefont {Heim}},
  \bibinfo {author} {\bibfnamefont {C.}~\bibnamefont {Peuntinger}}, \bibinfo
  {author} {\bibfnamefont {C.}~\bibnamefont {Wittmann}}, \bibinfo {author}
  {\bibfnamefont {C.}~\bibnamefont {Marquardt}}, \bibinfo {author}
  {\bibfnamefont {G.}~\bibnamefont {Leuchs}}, \ and\ \bibinfo {author}
  {\bibfnamefont {R.}~\bibnamefont {Filip}},\ }\href
  {http://stacks.iop.org/1367-2630/14/i=9/a=093048} {\bibfield  {journal}
  {\bibinfo  {journal} {New Journal of Physics}\ }\textbf {\bibinfo {volume}
  {14}},\ \bibinfo {pages} {093048} (\bibinfo {year} {2012})}\BibitemShut
  {NoStop}%
\bibitem [{\citenamefont {Vasylyev}\ \emph {et~al.}(2016)\citenamefont
  {Vasylyev}, \citenamefont {Semenov},\ and\ \citenamefont
  {Vogel}}]{Vogel2016PRL}%
  \BibitemOpen
  \bibfield  {author} {\bibinfo {author} {\bibfnamefont {D.}~\bibnamefont
  {Vasylyev}}, \bibinfo {author} {\bibfnamefont {A.~A.}\ \bibnamefont
  {Semenov}}, \ and\ \bibinfo {author} {\bibfnamefont {W.}~\bibnamefont
  {Vogel}},\ }\href {\doibase 10.1103/PhysRevLett.117.090501} {\bibfield
  {journal} {\bibinfo  {journal} {Phys. Rev. Lett.}\ }\textbf {\bibinfo
  {volume} {117}},\ \bibinfo {pages} {090501} (\bibinfo {year}
  {2016})}\BibitemShut {NoStop}%
\bibitem [{\citenamefont {Bohmann}\ \emph {et~al.}(2017)\citenamefont
  {Bohmann}, \citenamefont {Kruse}, \citenamefont {Sperling}, \citenamefont
  {Silberhorn},\ and\ \citenamefont {Vogel}}]{ChannelProbing}%
  \BibitemOpen
  \bibfield  {author} {\bibinfo {author} {\bibfnamefont {M.}~\bibnamefont
  {Bohmann}}, \bibinfo {author} {\bibfnamefont {R.}~\bibnamefont {Kruse}},
  \bibinfo {author} {\bibfnamefont {J.}~\bibnamefont {Sperling}}, \bibinfo
  {author} {\bibfnamefont {C.}~\bibnamefont {Silberhorn}}, \ and\ \bibinfo
  {author} {\bibfnamefont {W.}~\bibnamefont {Vogel}},\ }\href {\doibase
  10.1103/PhysRevA.95.063801} {\bibfield  {journal} {\bibinfo  {journal} {Phys.
  Rev. A}\ }\textbf {\bibinfo {volume} {95}},\ \bibinfo {pages} {063801}
  (\bibinfo {year} {2017})}\BibitemShut {NoStop}%
\bibitem [{\citenamefont {Residori}\ \emph {et~al.}(2017)\citenamefont
  {Residori}, \citenamefont {Onorato}, \citenamefont {Bortolozzo},\ and\
  \citenamefont {Arecchi}}]{Arecchi2016}%
  \BibitemOpen
  \bibfield  {author} {\bibinfo {author} {\bibfnamefont {S.}~\bibnamefont
  {Residori}}, \bibinfo {author} {\bibfnamefont {M.}~\bibnamefont {Onorato}},
  \bibinfo {author} {\bibfnamefont {U.}~\bibnamefont {Bortolozzo}}, \ and\
  \bibinfo {author} {\bibfnamefont {F.~T.}\ \bibnamefont {Arecchi}},\ }\href
  {\doibase 10.1080/00107514.2016.1243351} {\bibfield  {journal} {\bibinfo
  {journal} {Contemporary Physics}\ }\textbf {\bibinfo {volume} {58}},\
  \bibinfo {pages} {53} (\bibinfo {year} {2017})}\BibitemShut {NoStop}%
\bibitem [{\citenamefont {Mandel}(1958)}]{MandelFormula}%
  \BibitemOpen
  \bibfield  {author} {\bibinfo {author} {\bibfnamefont {L.}~\bibnamefont
  {Mandel}},\ }\href {http://stacks.iop.org/0370-1328/72/i=6/a=312} {\bibfield
  {journal} {\bibinfo  {journal} {Proceedings of the Physical Society}\
  }\textbf {\bibinfo {volume} {72}},\ \bibinfo {pages} {1037} (\bibinfo {year}
  {1958})}\BibitemShut {NoStop}%
\bibitem [{\citenamefont {Diament}\ and\ \citenamefont
  {Teich}(1970)}]{TeichModulation}%
  \BibitemOpen
  \bibfield  {author} {\bibinfo {author} {\bibfnamefont {P.}~\bibnamefont
  {Diament}}\ and\ \bibinfo {author} {\bibfnamefont {M.~C.}\ \bibnamefont
  {Teich}},\ }\href {\doibase 10.1364/JOSA.60.000682} {\bibfield  {journal}
  {\bibinfo  {journal} {J. Opt. Soc. Am.}\ }\textbf {\bibinfo {volume} {60}},\
  \bibinfo {pages} {682} (\bibinfo {year} {1970})}\BibitemShut {NoStop}%
\bibitem [{\citenamefont {Martienssen}\ and\ \citenamefont
  {Spiller}(1964)}]{GrainDisc}%
  \BibitemOpen
  \bibfield  {author} {\bibinfo {author} {\bibfnamefont {W.}~\bibnamefont
  {Martienssen}}\ and\ \bibinfo {author} {\bibfnamefont {E.}~\bibnamefont
  {Spiller}},\ }\href {\doibase 10.1119/1.1970023} {\bibfield  {journal}
  {\bibinfo  {journal} {American Journal of Physics}\ }\textbf {\bibinfo
  {volume} {32}},\ \bibinfo {pages} {919} (\bibinfo {year} {1964})}\BibitemShut
  {NoStop}%
\bibitem [{\citenamefont {Mehringer}\ \emph {et~al.}(2017)\citenamefont
  {Mehringer}, \citenamefont {Oppel},\ and\ \citenamefont {von
  Zanthier}}]{MMthermal}%
  \BibitemOpen
  \bibfield  {author} {\bibinfo {author} {\bibfnamefont {T.}~\bibnamefont
  {Mehringer}}, \bibinfo {author} {\bibfnamefont {S.}~\bibnamefont {Oppel}}, \
  and\ \bibinfo {author} {\bibfnamefont {J.}~\bibnamefont {von Zanthier}},\
  }\href {\doibase 10.1007/s00340-017-6775-y} {\bibfield  {journal} {\bibinfo
  {journal} {Applied Physics B}\ }\textbf {\bibinfo {volume} {123}},\ \bibinfo
  {pages} {200} (\bibinfo {year} {2017})}\BibitemShut {NoStop}%
\bibitem [{\citenamefont {Kondakci}\ \emph {et~al.}(2015)\citenamefont
  {Kondakci}, \citenamefont {Abouraddy},\ and\ \citenamefont
  {Saleh}}]{SalehNatPhys2015}%
  \BibitemOpen
  \bibfield  {author} {\bibinfo {author} {\bibfnamefont {H.~E.}\ \bibnamefont
  {Kondakci}}, \bibinfo {author} {\bibfnamefont {A.~F.}\ \bibnamefont
  {Abouraddy}}, \ and\ \bibinfo {author} {\bibfnamefont {B.~E.~A.}\
  \bibnamefont {Saleh}},\ }\href {http://dx.doi.org/10.1038/nphys3482}
  {\bibfield  {journal} {\bibinfo  {journal} {Nature Physics}\ }\textbf
  {\bibinfo {volume} {11}},\ \bibinfo {pages} {930} (\bibinfo {year}
  {2015})}\BibitemShut {NoStop}%
\bibitem [{\citenamefont {Kondakci}\ \emph {et~al.}(2016)\citenamefont
  {Kondakci}, \citenamefont {Szameit}, \citenamefont {Abouraddy}, \citenamefont
  {Christodoulides},\ and\ \citenamefont {Saleh}}]{SalehOptica2016}%
  \BibitemOpen
  \bibfield  {author} {\bibinfo {author} {\bibfnamefont {H.~E.}\ \bibnamefont
  {Kondakci}}, \bibinfo {author} {\bibfnamefont {A.}~\bibnamefont {Szameit}},
  \bibinfo {author} {\bibfnamefont {A.~F.}\ \bibnamefont {Abouraddy}}, \bibinfo
  {author} {\bibfnamefont {D.~N.}\ \bibnamefont {Christodoulides}}, \ and\
  \bibinfo {author} {\bibfnamefont {B.~E.~A.}\ \bibnamefont {Saleh}},\ }\href
  {\doibase 10.1364/OPTICA.3.000477} {\bibfield  {journal} {\bibinfo  {journal}
  {Optica}\ }\textbf {\bibinfo {volume} {3}},\ \bibinfo {pages} {477} (\bibinfo
  {year} {2016})}\BibitemShut {NoStop}%
\bibitem [{\citenamefont {B\'{e}dard}(1967)}]{BedardInverse}%
  \BibitemOpen
  \bibfield  {author} {\bibinfo {author} {\bibfnamefont {G.}~\bibnamefont
  {B\'{e}dard}},\ }\href {\doibase 10.1364/JOSA.57.001201} {\bibfield
  {journal} {\bibinfo  {journal} {J. Opt. Soc. Am.}\ }\textbf {\bibinfo
  {volume} {57}},\ \bibinfo {pages} {1201} (\bibinfo {year}
  {1967})}\BibitemShut {NoStop}%
\bibitem [{\citenamefont {Pe{\v{r}}ina}\ and\ \citenamefont
  {Mi{\v{s}}ta}(1969)}]{Perina1969}%
  \BibitemOpen
  \bibfield  {author} {\bibinfo {author} {\bibfnamefont {J.}~\bibnamefont
  {Pe{\v{r}}ina}}\ and\ \bibinfo {author} {\bibfnamefont {L.}~\bibnamefont
  {Mi{\v{s}}ta}},\ }\href {\doibase 10.1002/andp.19694770707} {\bibfield
  {journal} {\bibinfo  {journal} {Annalen der Physik}\ }\textbf {\bibinfo
  {volume} {477}},\ \bibinfo {pages} {372} (\bibinfo {year}
  {1969})}\BibitemShut {NoStop}%
\bibitem [{\citenamefont {Byrne}\ \emph {et~al.}(1993)\citenamefont {Byrne},
  \citenamefont {Haughton},\ and\ \citenamefont {Jiang}}]{ByrneInverse}%
  \BibitemOpen
  \bibfield  {author} {\bibinfo {author} {\bibfnamefont {C.}~\bibnamefont
  {Byrne}}, \bibinfo {author} {\bibfnamefont {D.}~\bibnamefont {Haughton}}, \
  and\ \bibinfo {author} {\bibfnamefont {T.}~\bibnamefont {Jiang}},\ }\href
  {http://stacks.iop.org/0266-5611/9/i=1/a=002} {\bibfield  {journal} {\bibinfo
   {journal} {Inverse Problems}\ }\textbf {\bibinfo {volume} {9}},\ \bibinfo
  {pages} {39} (\bibinfo {year} {1993})}\BibitemShut {NoStop}%
\bibitem [{\citenamefont {Earnshaw}\ and\ \citenamefont
  {Haughey}(1996)}]{SplinesInverse}%
  \BibitemOpen
  \bibfield  {author} {\bibinfo {author} {\bibfnamefont {J.~C.}\ \bibnamefont
  {Earnshaw}}\ and\ \bibinfo {author} {\bibfnamefont {D.}~\bibnamefont
  {Haughey}},\ }\href {\doibase 10.1063/1.1147540} {\bibfield  {journal}
  {\bibinfo  {journal} {Review of Scientific Instruments}\ }\textbf {\bibinfo
  {volume} {67}},\ \bibinfo {pages} {4387} (\bibinfo {year}
  {1996})}\BibitemShut {NoStop}%
\bibitem [{\citenamefont {Lawson}\ and\ \citenamefont {Hanson}(1987)}]{NNLS}%
  \BibitemOpen
  \bibfield  {author} {\bibinfo {author} {\bibfnamefont {C.~L.}\ \bibnamefont
  {Lawson}}\ and\ \bibinfo {author} {\bibfnamefont {R.~J.}\ \bibnamefont
  {Hanson}},\ }\href@noop {} {\emph {\bibinfo {title} {Solving Least Squares
  Problems}}}\ (\bibinfo  {publisher} {SIAM},\ \bibinfo {year}
  {1987})\BibitemShut {NoStop}%
\bibitem [{\citenamefont {Ho{\v{s}}{\'a}k}\ and\ \citenamefont
  {Je{\v{z}}ek}(2018)}]{Hosak2018}%
  \BibitemOpen
  \bibfield  {author} {\bibinfo {author} {\bibfnamefont {R.}~\bibnamefont
  {Ho{\v{s}}{\'a}k}}\ and\ \bibinfo {author} {\bibfnamefont {M.}~\bibnamefont
  {Je{\v{z}}ek}},\ }\href {https://arxiv.org/abs/1801.02433} {\bibfield
  {journal} {\bibinfo  {journal} {arXiv:1801{.}02433}\ } (\bibinfo {year}
  {2018})}\BibitemShut {NoStop}%
\end{thebibliography}%

\end{document}